\documentclass{appolb}
\usepackage{epsfig}

\newcommand{\pom}{{I\!\!P}}
\newcommand{\apom}{\mbox{$\alpha_{\pom}$}}

\newcommand{\qqbar}{\ensuremath{q\bar{q}}}
\newcommand{\jpsi}{\ensuremath{J/\psi}}

\newcommand{\ttra}{\ensuremath{|t|}}
\newcommand{\qsq}{\ensuremath{Q^2}}

\newcommand{\gevsq}{\mbox{${\rm GeV}^2$}}

\newcommand{\siggp}{\ensuremath{\sigma(\gamma^{*} p \rightarrow \rho p)}}
\newcommand{\rfour}{\mbox{$r^{04}_{00}$}}

\newcommand{\rfivecomb}{\mbox{$r^5_{00} + 2 r^5_{11}$}}
\newcommand{\ronecomb}{\mbox{$r^1_{00} + 2 r^1_{11}$}}

\begin{document}
\title{Diffractive $\rho$ Meson Electroproduction \\
           at High $Q^2$ and High $|t|$%
}
\author{Xavier Janssen
        \thanks{The author is supported by the F.R.I.A (Belgium)}, \\ 
        on behalf of the H1 collaboration
\address{Universit\'e Libre de Bruxelles, Bd du Triomphe, CP 230, \\
         B-1050 Brussels, Belgium. \\
         e-mail: janssen@hep.iihe.ac.be }
}
\maketitle
\begin{abstract}
The electroproduction of $\rho$ mesons is studied at HERA with the H1 detector 
at high $Q^2$ and high \ttra. Cross sections are measured as a 
function of $Q^2$, $W$ and $t$. The $W$ dependence of the $\gamma^{*} p$ 
cross section 
is observed to increase with $Q^2$ from values compatible with soft
Pomeron exchange at low $Q^2$ to a hard dependence at large $Q^2$. Spin
density matrix elements are measured and their dependence is compared 
with a two gluon exchange model.
\end{abstract}
\PACS{PACS numbers come here}

\section{Introduction} 

We present results on the diffractive electroproduction of $\rho$ mesons
in $ep$ scattering at high energy: $e p \rightarrow e \rho Y$, 
$\rho \rightarrow \pi^+ \pi^-$, where $Y$ is either a proton (``elastic''
scattering) or a baryonic system of mass $M_Y$ much lower than the 
$\gamma^* p$ center of mass (cms) energy $W$ 
(``proton dissociative'' scattering.)  

At low $Q^2$, the negative of the intermediate photon four-momentum squared,
the $\gamma^* p$ cross section is characterized by a ``soft''
energy dependence due to  pomeron ($\pom$) exchange with 
$ {\rm d}\sigma / {\rm d}t \propto W^{4(\apom(t)-1)}$, where the soft
pomeron trajectory is parametrized~\cite{dl} as
$\apom(t) =\apom(0) - \alpha' \cdot \ttra \simeq 1.08 - 0.25 \cdot \ttra$, 
$t$ being the square of the four-momentum transfer to the proton. 

For \qsq\ larger than a few \gevsq, perturbative QCD (pQCD) is expected to
apply and diffractive $\rho$ production is viewed in the proton rest
frame as a sequence of three processes well separated in time : the photon
fluctuation into a \qqbar\ pair, the hard interaction of the \qqbar\ pair
with the proton via the exchange of two gluons in a color-singlet state,
and the the \qqbar\ pair recombination into a real $\rho$ meson.
The cross section is then
proportional to the square of the gluon density in the proton,
which provides a fast increase of the $\gamma^{*} p$ cross section with
the energy (``hard'' behavior)~\cite{ryskin}.


The data, taken with the H1 detector in 1997 and 2000, correspond
to integrated luminosities of 6 ${\rm pb}^{-1}$ and 42 ${\rm pb}^{-1}$,
respectively. The 1997
data are dedicated to the study of the ``high \ttra'' regime~\cite{h1:rho97}
in the kinematic domain $\ttra < 3 $ \gevsq, $2.5 < \qsq < 30$ \gevsq\ and
$40 < W < 120$ GeV. This sample includes both elastic and proton dissociative
events. The ``high \qsq'' regime is investigated using the 2000 data
in the kinematic domain $2.5 < \qsq < 60$ \gevsq,
$40 < W < 180$ GeV and  $\ttra < 0.5 $ \gevsq\ (elastic channel only).

\section{Cross sections} 

\begin{figure}[htb]
 \begin{center}
   \epsfig{file=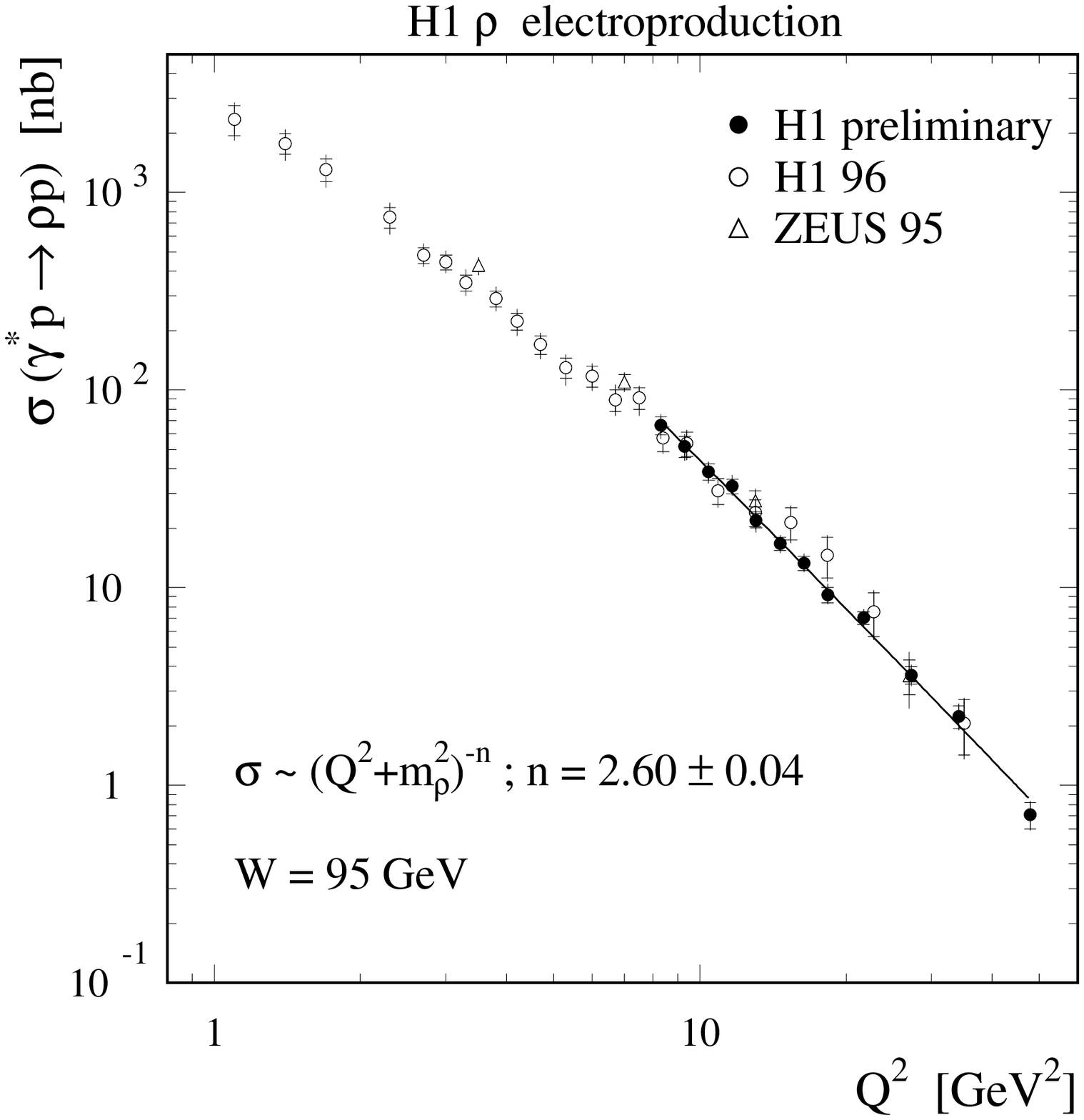,width=0.48\textwidth}
   \epsfig{file=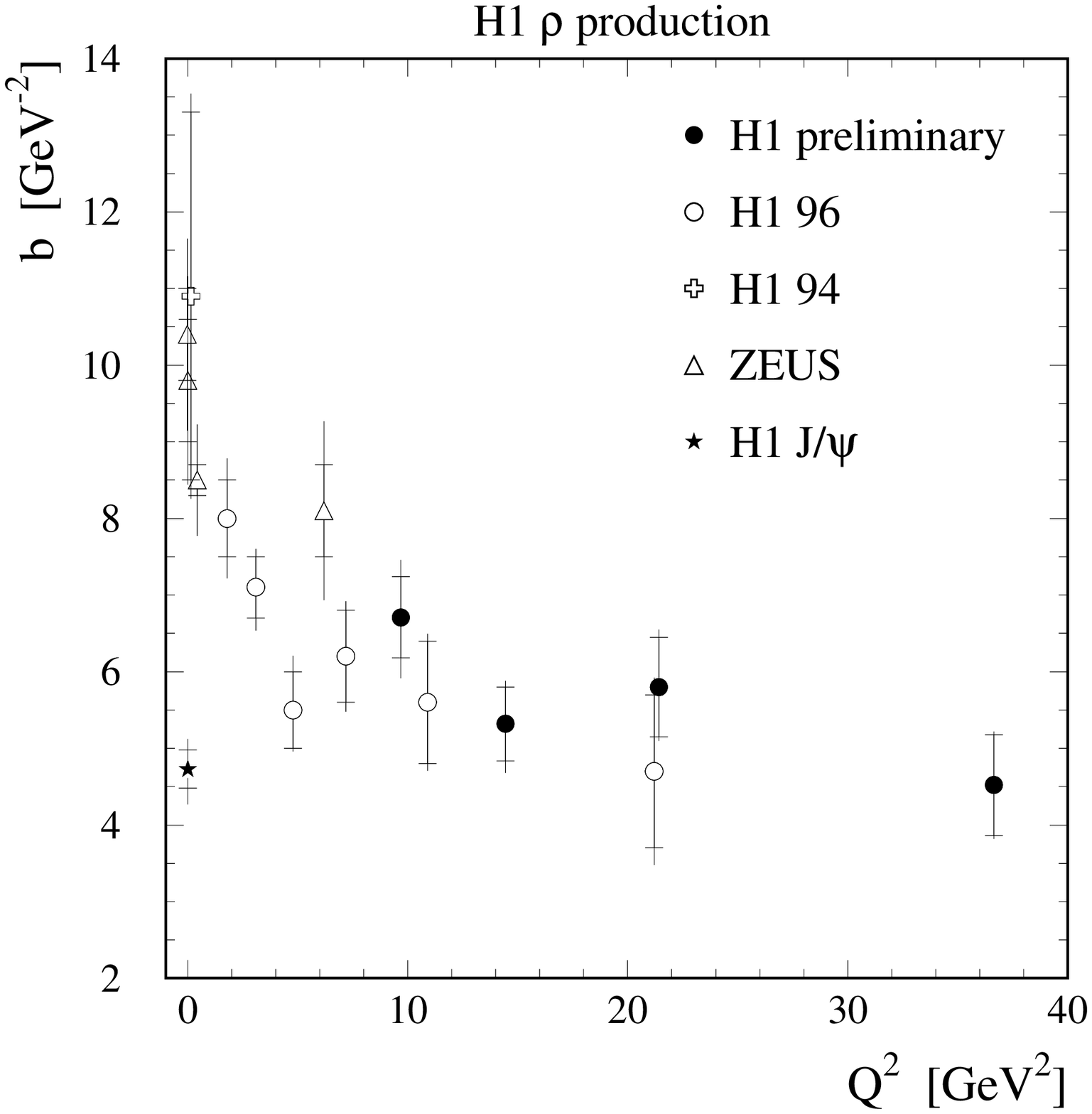,width=0.48\textwidth}
 \end{center}
 \vspace{-0.7cm}
 \setlength{\unitlength}{1.0cm}
 \begin{picture}(12.6,0.1)
    \put(3.15,0.0){(a)}
    \put(9.45,0.0){(b)}
 \end{picture}
 \vspace{-0.3cm}
 \caption{(a) \siggp\ as a function of \qsq\ for $W = 95$ GeV, 
              together with previous measurements~\cite{h1:rho96,zeus:rho95}. 
              Full line: see text.
          (b) Results of fits to exponential distributions of the $t$
              dependence of \siggp,
              presented as a function of \qsq\ together with previous
              for $\rho$~\cite{h1:rho96,zeus:rho95,h1:rho94gp} and
              \jpsi~\cite{h1:jpsi} measurements.}
 \label{fig:q2t}
 \vspace{-0.2cm}
\end{figure}

The \qsq\ dependence of the $\gamma^{*} p \rightarrow \rho p$
cross section for $W = 95$ GeV is presented in Fig.~\ref{fig:q2t}a.
A fit of the form 
${\rm d} \sigma / {\rm d} Q^2 \propto 1 / (Q^2  + m_\rho^2)^{n}$
for the range $8 < Q^2 < 60$~\gevsq\ results in $n = 2.60 \pm 0.04$
(full line).

The $t$ dependence of the cross section is parametrized as 
${\rm d} \sigma / {\rm d} t \propto exp(-b \ttra)$. 
Measured values of the $b$ parameter are
shown as a function of \qsq\ in Fig.~\ref{fig:q2t}b: 
$b$ decreases as \qsq\ increases, reflecting the decrease of the 
transverse size of the \qqbar\ pair, down to values 
close to the \jpsi\ case in photoproduction~\cite{h1:jpsi}. 

\begin{figure}[htb]
 \begin{center}
   \epsfig{file=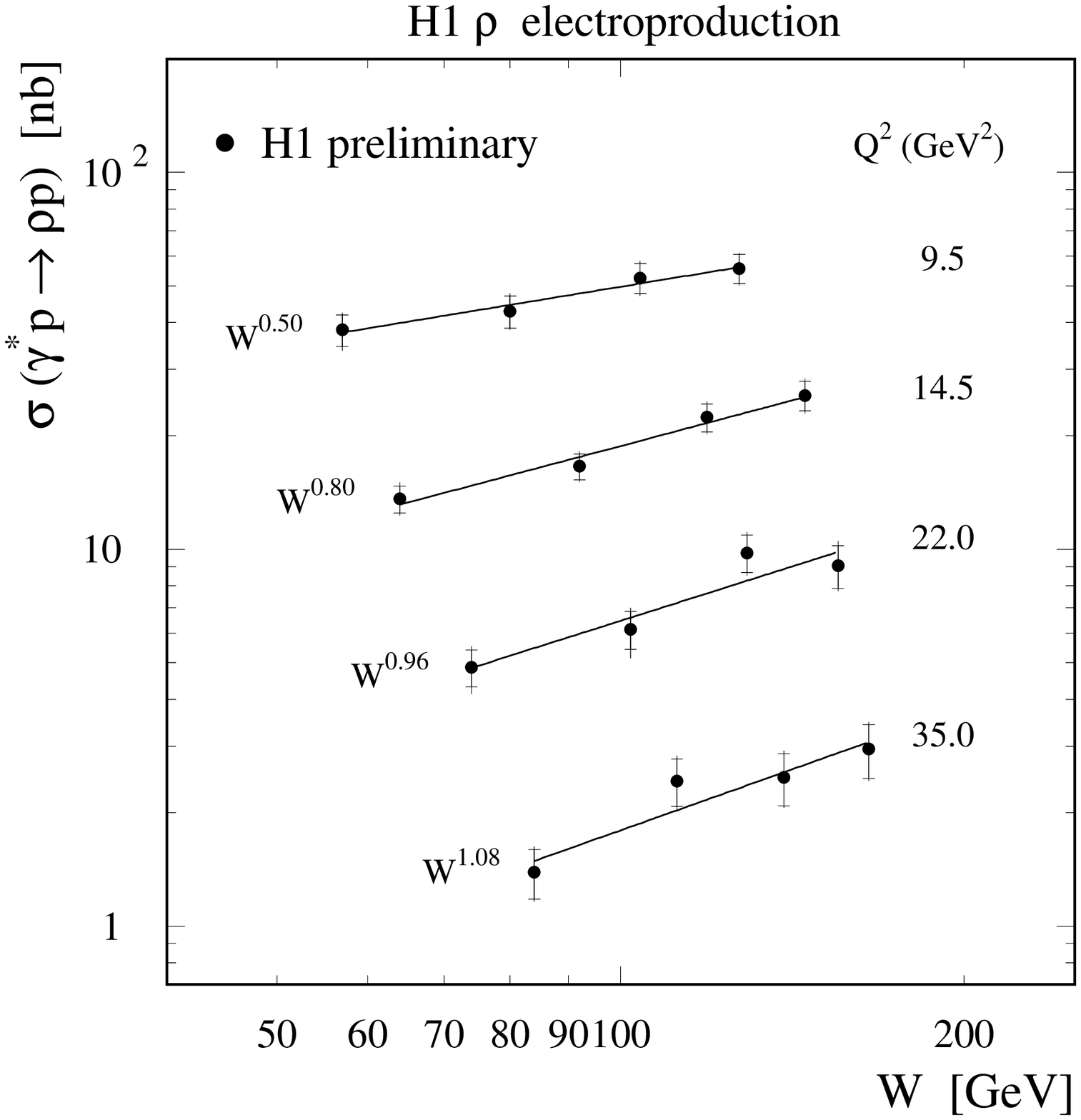,width=0.48\textwidth}
   \epsfig{file=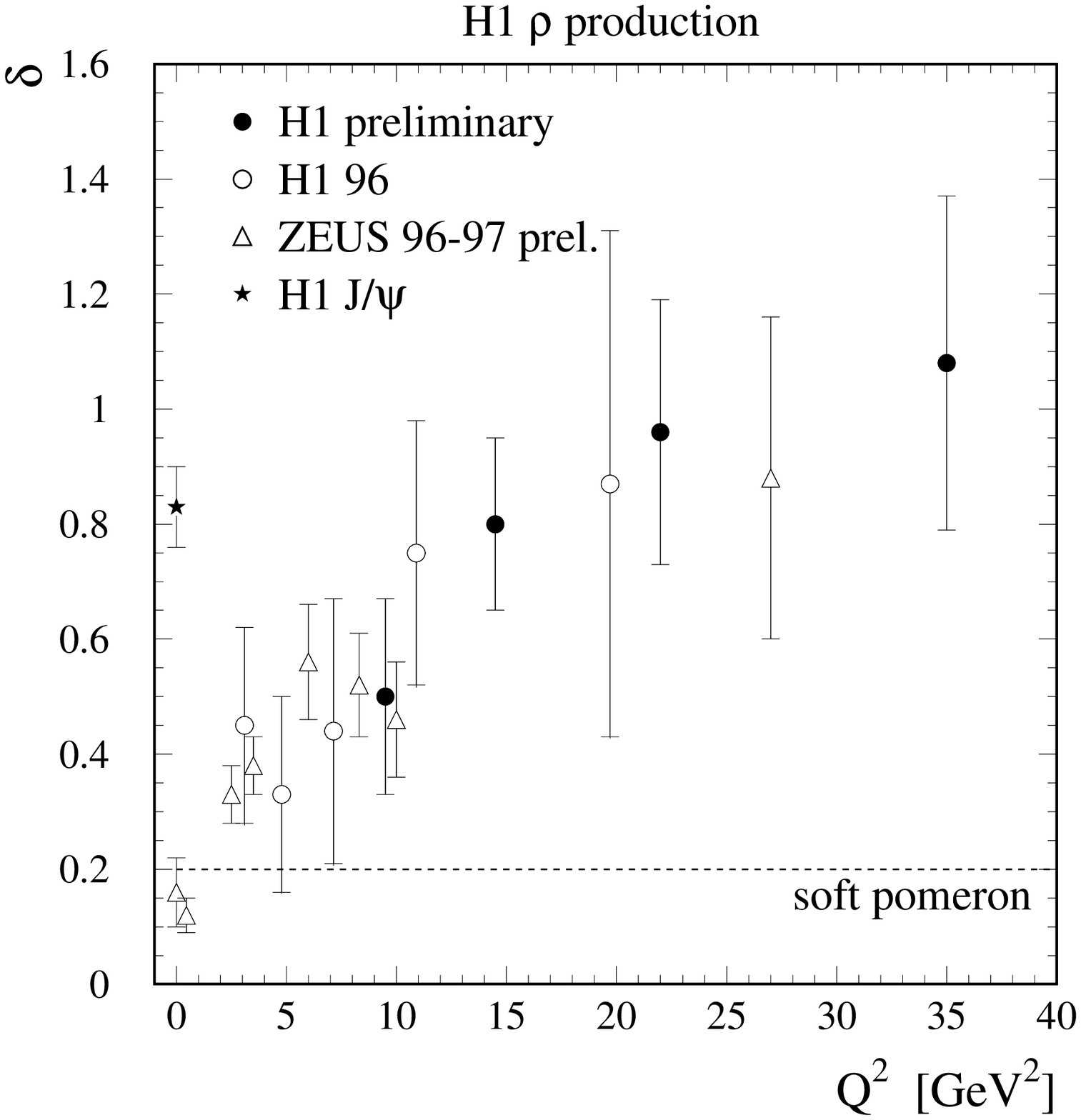,width=0.48\textwidth}
 \end{center}
 \vspace{-0.7cm}
 \setlength{\unitlength}{1.0cm}
 \begin{picture}(12.6,0.1)
    \put(3.15,0.0){(a)}
    \put(9.45,0.0){(b)}
 \end{picture}
 \caption{(a) \siggp\ as a function of $W$ in four \qsq\ intervals.
          (b) Results of fits of the form 
              $W^\delta$ to the $W$ dependence of \siggp\ (full lines
              in (a)), presented as a function of \qsq\ together with
              previous $\rho$~\cite{h1:rho96,zeus:rho9697} and
              \jpsi\ \cite{h1:jpsi} measurements.}  
 \label{fig:w}
 \vspace{-0.2cm}
\end{figure}

The $W$ dependence of the $\gamma^{*} p \rightarrow \rho p$
cross section, parametrised
as $ \sigma (W) \propto W^\delta$, has been measured in 
four \qsq\ intervals (Fig.~\ref{fig:w}).
A clear transition from a ``soft'' to a ``hard'' behavior is observed
when \qsq\ increases, reaching values similar to the \jpsi\ in
photoproduction~\cite{h1:jpsi}, where the charm quark mass acts 
as a hard scale in pQCD.

\section{Spin density matrix elements} 

The measurement of the production and decay angular distibutions gives 
access to the spin density matrix elements, which characterise the helicity
states of the photon and the $\rho$ meson, and which are bilinear
combinations of the helicity amplitudes 
$T_{\lambda_{\rho} \lambda_{\gamma}}$~\cite{sw},
where $\lambda_{\rho}$,$\lambda_{\gamma}=-1,0,1$ stand for the helicities of
the $\rho$ meson and the photon, respectively. The matrix element \rfour\
and the combinations \rfivecomb\ and \ronecomb\ have been measured 
for the ``high \ttra'' and the ``high \qsq'' data samples.

In case of $s$-channel helicity conservation (SCHC), the helicity of the
$\rho$ meson and the photon are the same and the combinations \rfivecomb\
and \ronecomb\ vanish. Calculations based on pQCD~\cite{ivanov,nikolaev} 
predict SCHC violations with the following qualitative $t$ dependences:  
a ratio constant with $t$ for the helicity conserving amplitudes
($T_{00}$ and $T_{11}$),
a $\sqrt {|t|}$ dependence for the ratio of the single
helicity flip ($T_{01}$ and $T_{10}$) to the non-flip amplitudes, 
a linear dependence for the ratio of the double flip ($T_{1-1}$) to the
non-flip amplitudes. 
The matrix element \rfour\ is related to the ratio of the longitudinal
to transverse non-flip amplitudes; the $t$ dependence of the \rfivecomb\
combination is dominated by the $T_{01}$ amplitude, and that of the 
\ronecomb\ combination by $|T_{01}|^2$.

\begin{figure}[htb]
 \begin{center}
   \epsfig{file=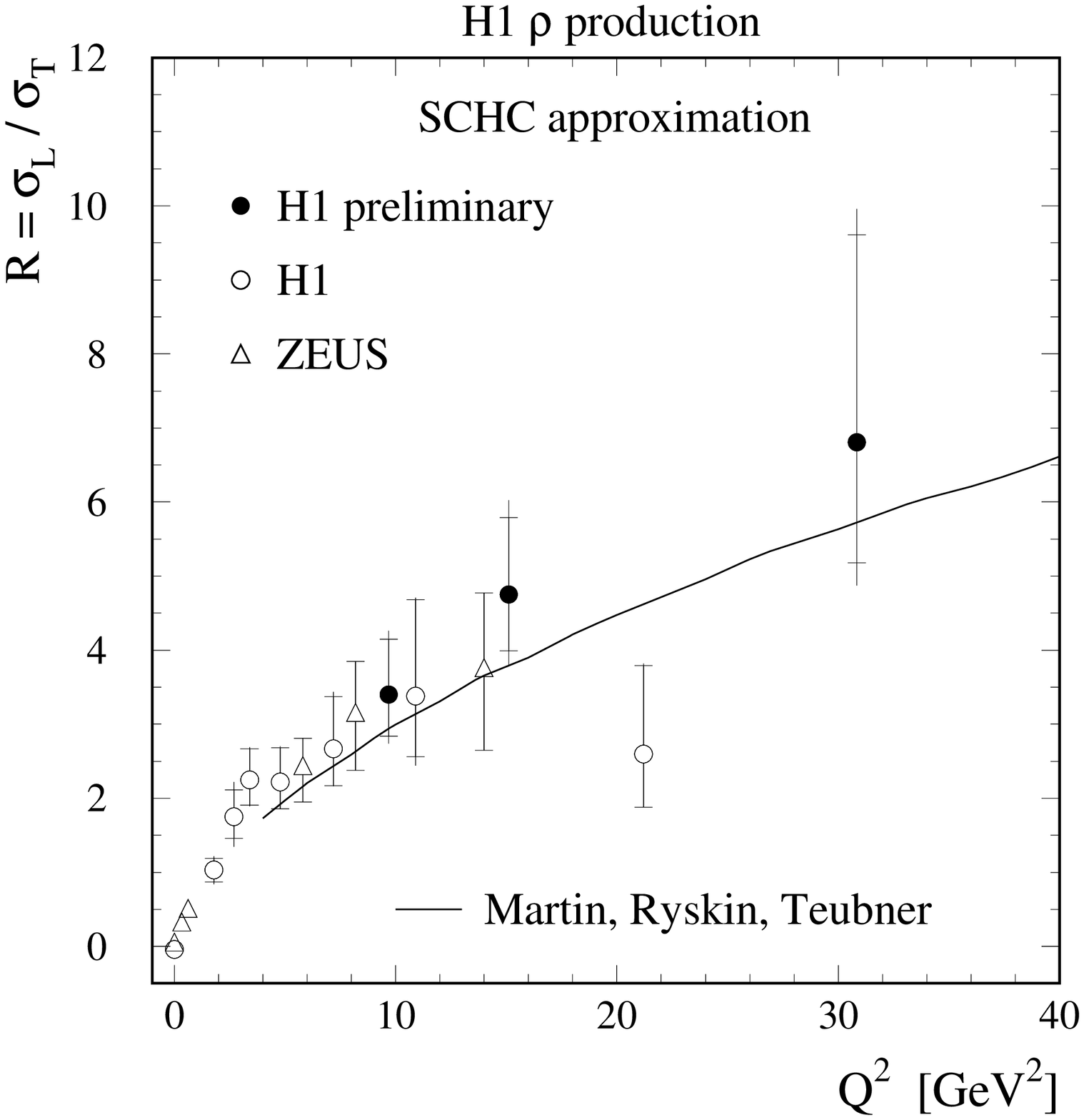,width=0.48\textwidth}
   \epsfig{file=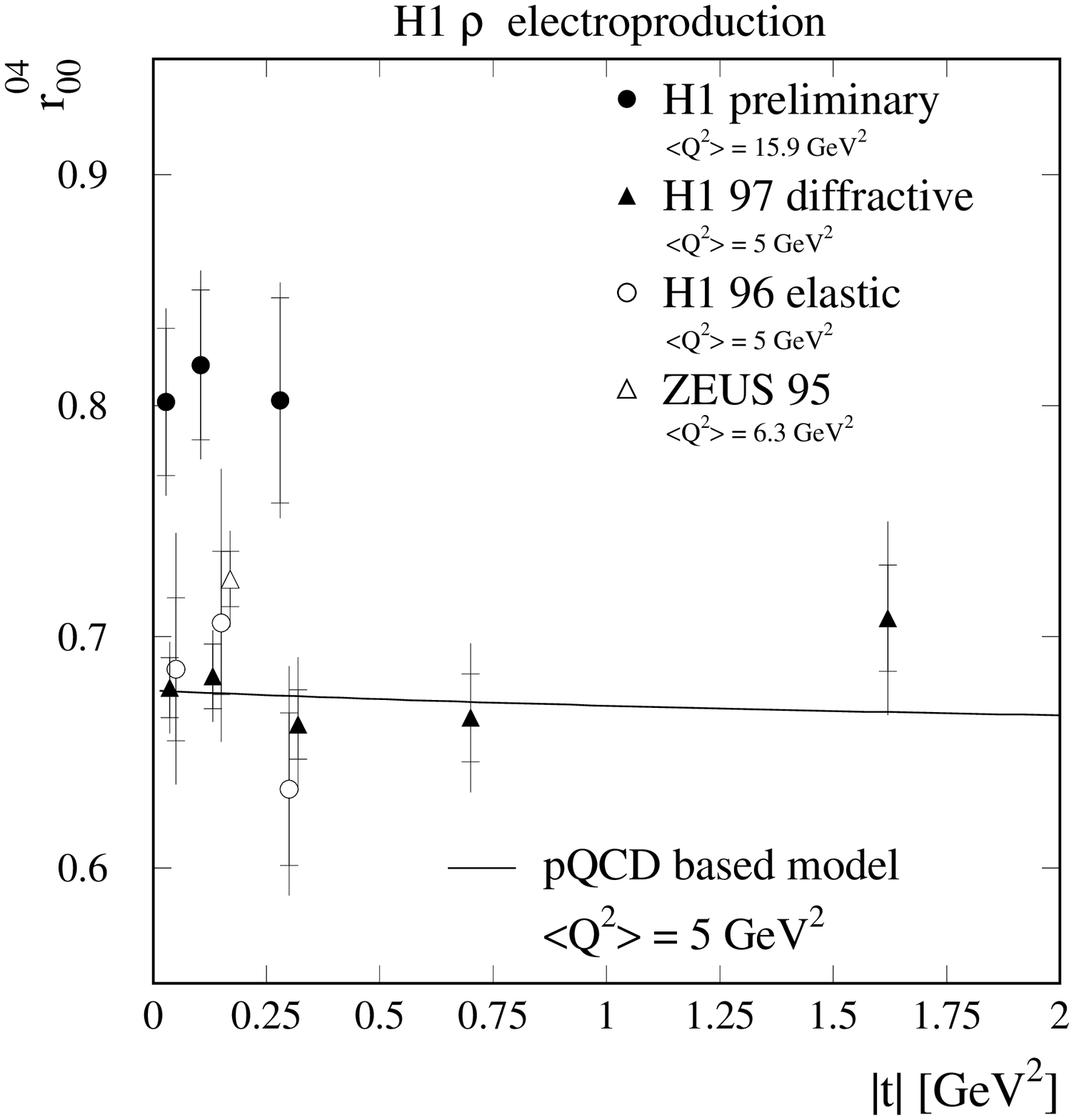,width=0.48\textwidth}
 \end{center}
 \vspace{-0.7cm}
 \setlength{\unitlength}{1.0cm}
 \begin{picture}(12.6,0.1)
    \put(3.15,0.0){(a)}
    \put(9.45,0.0){(b)}
 \end{picture}
 \caption{Measurements of 
          (a) $R = \sigma_L / \sigma_T$ as a function of \qsq,
              together with previous 
              measuments~\cite{h1:rho96,h1:rho94gp,zeus:rho95,zeus:rhogp}
          (b) \rfour\  as a function of \ttra, together with previous
              measuments~\cite{h1:rho96,zeus:sdme}
              Full lines: see text.}
 \label{fig:r4}
 \vspace{-0.2cm}
\end{figure}

Fig.~\ref{fig:r4}a presents  
$R = \sigma_L / \sigma_T$ as a function of \qsq, extracted
from the measurement of the \rfour\ 
matrix element. The significant increase of $R$ with \qsq\ 
is compatible the pQCD based prediction~\cite{mrt}, shown as a full line.

\begin{figure}[htb]
 \begin{center}
   \epsfig{file=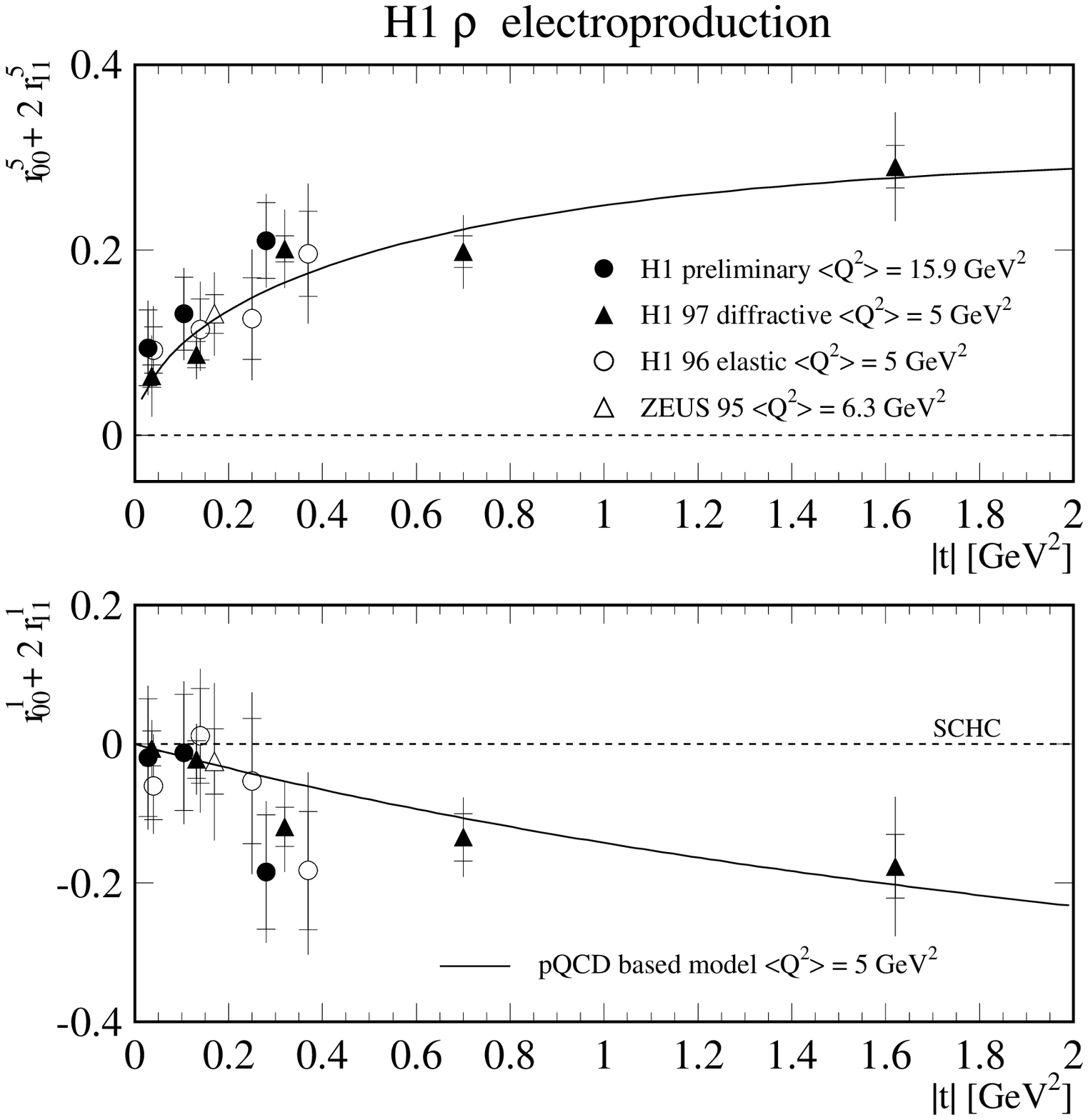,width=0.7\textwidth}
 \end{center}
 \vspace{-0.9cm}
 \setlength{\unitlength}{1.0cm}
 \begin{picture}(.1,0.1)
    \put(1.0,6.2){(a)}
    \put(1.0,2.5){(b)}
 \end{picture}
 \vspace{-0.5cm}
 \caption{Measurements of (a) \rfivecomb\ and (b) \ronecomb\ as a 
          function of \ttra,  together with previous
          measuments~\cite{h1:rho96,zeus:sdme}.
          Full lines: see text }
 \label{fig:r51t}
 \vspace{-0.4cm}
\end{figure}

The $t$ dependence of the \rfour\ matrix element 
and of the combinations \rfivecomb\ and \ronecomb\ are presented in
Figs.~\ref{fig:r4}b and~\ref{fig:r51t}. The expected SCHC violation 
is confirmed. The ``high \ttra'' sample is well described by 
prediction based on the model from ref.~\cite{ivanov}. 
The very weak dependence of \rfour\ with 
\ttra\ indicates that, in the measured range, the longitudinal and
the transverse cross sections have similar $t$ dependences. 

\section{Conclusions}

The electroproduction of $\rho$ mesons has been studied at HERA with the
H1 detector at high \ttra\ and high \qsq. 
The W dependence of the $\gamma^{*} p$ cross section 
is observed to increase with $Q^2$ from values compatible with soft
Pomeron exchange at low $Q^2$ to a hard dependence at large $Q^2$.
At high \qsq,
the $W$ and the $t$ dependences for $\rho$
electroproduction become similar to \jpsi\ 
photoproduction. The \rfour\ spin density matrix elements and the
combinations \rfivecomb\ and \ronecomb\ have been measured as a function
of $t$. No significant $t$ dependence for \rfour, ant the $t$ 
dependences of the \rfivecomb\ and \ronecomb\ combinations are
in agreement with perturbative QCD predictions.

\vspace{-0.5cm}


\end{document}